\documentclass[global,twocolumn]{svjour}
\usepackage[dvips]{graphics}
\usepackage[intlimits]{amsmath}
\usepackage{graphicx}
\usepackage[english]{babel}
\usepackage{epsfig}
\usepackage{amssymb}
\usepackage{psfrag}
\graphicspath{{../FIGS}{FIGS/}}
\journalname{Applied Physics B}

\begin{document}

\title{Real-time phase-shift detection of the surface plasmon resonance}

\author{T. K\"{o}nig\inst{1} \and M. Weidem\"{u}ller\inst{2} \and A. Hemmerich\inst{1}}
\institute{Institut f\"{u}r Laser-Physik, Universit\"{a}t Hamburg, Luruper Chaussee 149, 22761 Hamburg, Germany \and Physikalisches Institut, Universit\"{a}t Freiburg, Hermann-Herder-Stra\ss e 3, 79104 Freiburg, Germany}

\date{Received: date / Revised version: date}

\maketitle

\begin{abstract}
We investigate a method to directly measure the phase of a laser beam reflected from a metallic film after excitation of surface plasmon polaritons. This method permits real time access to the phase information, it increases the possible speed of data acquisition, and it may thus prove useful for increasing the sensitivity of surface plasmon based sensors. 
\end{abstract}


\section{Introduction}
\label{sec:intro}
Sensors based upon laser-aided detection of the surface plasmon resonance (SPR) have become a widely used tool in a variety of chemical and biological applications \cite{Hom:99} in which the index of refraction of a liquid or gaseous medium needs to be carefully monitored. In SPR sensors a laser beam is reflected from a thin metallic film with the angle of incidence and the thickness of the film adjusted such that a SPR can be excited. The properties of the reflected beam in the vicinity of the SPR display a sensitive dependence on the index of refraction in the medium adjacent to the metallic film. While in commercially available SPR sensors the intensity of the reflected beam near the SPR is analyzed, more involved scenarios have been recently investigated in order to utilize information on the phase of the reflected beam \cite{Kab:98,She:98,Wu:03,Wu:04}. Owing to the complex dielectric function of the metallic film this phase near the SPR displays an extremely steep dependence on the angle of incidence, if the film thickness is properly matched to the frequency of the incident light. This has raised expectations that phase sensitive detection could in principle improve the sensitivity of SPR sensors. Unfortunately, according to Ref.\cite{Ran:06} such hopes seem to be frustrated because the phase cannot be directly measured but has to be inferred from intensity measurements (e.g., using a phase-step algorithm \cite{Gre:92}), which is hampered by the fact that the reflected intensity becomes very small in the vicinity of the SPR. While this consideration applies for moderate laser intensities, when shot noise is the main noise source, in practical experiments phase sensitive detection schemes may nevertheless be very useful to increase the detection sensitivity without requiring expensive intensity measurement devices offering extreme dynamics and precision. In this work we point out a method to directly measure the phase of a laser beam reflected from a metallic film under excitation of a SPR. This method permits real time access to the phase and thus significantly increases the possible speed of data acquisition as compared to indirect phase detection techniques. In our experiments a sensitivity with respect to changes of the index of refraction adjacent to the metallic film of about $10^{-8}$ can be reached in 0.1 s observation time with standard detection equipment operating far from the shot noise level with moderate dynamics and precision. 

\section{Method of Phase Determination}\label{sec:phase}
\begin{figure}
\center
\resizebox{0.45\textwidth}{!}{\includegraphics{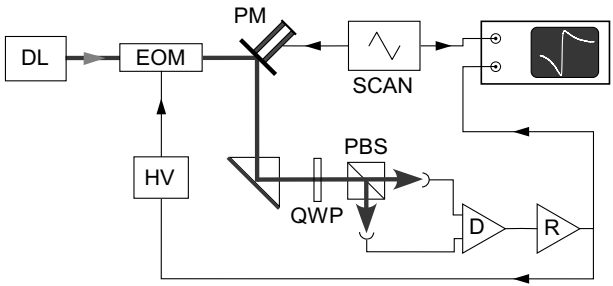}}
\caption{Experimental setup. DL = grating stabilized diode laser; EOM = electro optical modulator; PM = PZT tilted mirror; QWP = quarter wave plate; PBS = polarization beam splitter; D = differential amplifier; R = regulator circuit; HV = high voltage amplifier. }
\label{fig:setup}    
\end{figure}

We employ the Kretschmann-Raether configuration \cite{Kre:68} to excite the SPR using the optical scheme shown in Fig.\ref{fig:setup}. The light source is a single-frequency diode laser (used emission power $P = 6\,$mW) emitting at 780 nm with a few MHz bandwidth \cite{Ric:95}. Its linearly polarized electric field is adjusted to comprise equally large components $E_p$ parallel and $E_s$ perpendicular to the drawing plane. The light traverses an electro-optic modulator (EOM), which shifts $E_s$ by an adjustable phase $\phi_{\rm{EOM}}$. The $E_p$ component can excite the SPR with the corresponding complex amplitude reflection coefficient $r_p \equiv |r_p| e^{i \delta_p}$ while $E_s$ is unaffected by the plasmon and serves as a phase reference with its reflection coefficient assumed to be unity. After reflection by the metallic film (the angle of incidence being adjusted by a mirror mounted on a piezo-electric transducer (PZT)) the light beam traverses a quarter wave plate and a polarization beam splitter. The difference between the powers on both exits $\Delta P = 2 P |r_p| \sin(\delta_p - \phi_{\rm{EOM}})$ is used as the error input of a simple regulator circuit (consisting of an inverter and an integrator), which servo-controls $\phi_{\rm{EOM}}$ in order to keep  $\Delta P = 0$ and thus 
\begin{equation}
\phi_{\rm{EOM}} = \delta_p \, .	
\end{equation}
Accordingly, the phase of the reflected beam introduced by the SPR is simply obtained by monitoring the EOM control voltage with the servo loop closed.

\section{Experimental Considerations}\label{sec:exp}
\begin{figure}
\center
\resizebox{0.40\textwidth}{!}{
 \includegraphics{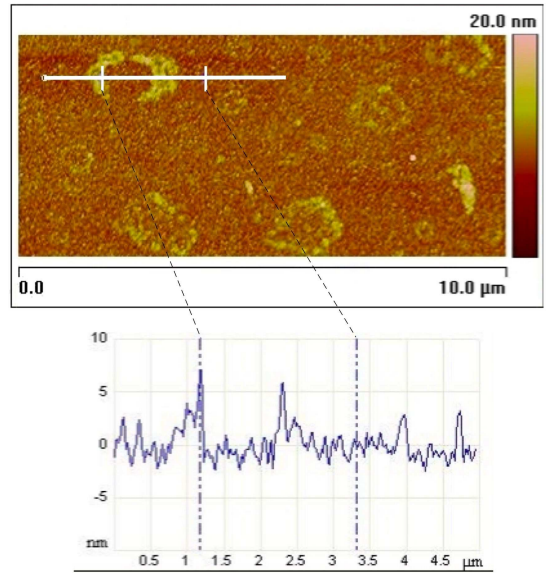}}
\caption{Atomic force microscopy picture of silver film, showing islands of increased roughness. The lower graph shows a section through one of these islands.}
\label{fig:afm}    
\end{figure}

For SPR excitation a thin silver film was evaporated on a microscope slide. Its surface roughness was determined with an atomic force microscope (see Fig. \ref{fig:afm}) to be $1\,$nm. After coating the sample maintained $1\,$ nm rms-roughness in most areas with some islands showing increased roughness up to $7\,$nm. The detail in Fig. \ref{fig:afm} shows a typical profile of a rough area. The coated slide was fixed with index matching liquid to the base of a $90^\circ$ prism, which facilitates the incoupling of the laser beam into the glass substrate. We thus obtain a stack of three dielectric layers, namely glass, metallic, and air with dielectric constants $\epsilon_{\rm{glass}}$, $\epsilon_{\rm{Ag}}$ and $\epsilon_{\rm{air}}$. The light enters the silver film through the glass, which provides the wave vector increase required to match the dispersion of the light and the surface plasmon. The incident laser beam is well collimated (beam waist = 1.6 mm) corresponding to a beam divergence of about $300 \, \mu$rad. The angle of incidence can be varied with the PZT-mirror within a range of 16 mrad with $0.1 \, \mu$rad resolution. The modulus $|r_p| = |r_p(\theta, d)|$ and the phase $\delta_p = \delta_p(\theta, d)$ of the amplitude reflection coefficient depend on the angle of incidence $\theta$ and the thickness of the metallic film $d$ and may be readily calculated from the Fresnel relations \cite{Rae:88}. All optical elements including the prism are positioned with standard opto-mechanical mounts, i.e., no specific measures were taken to reduce undesired angle fluctuations due to mechanical vibrations. The EOM lets us add a phase to the $E_s$ component of the incident beam of more than $2 \pi$ with a precision of $12\,$mrad determined by the noise of our high voltage source. Intensity detection is accomplished with standard photo diodes operating with $20\,$MHz bandwidth.
\begin{figure}
\center
\resizebox{0.40\textwidth}{!}{
\includegraphics{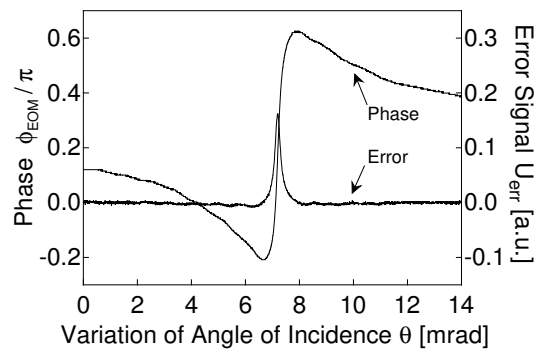}}
\caption{The error signal $U_{\rm{err}} \propto \Delta P$ at the exit of the differential amplifier (D) in Fig.\ref{fig:setup} and the voltage $\phi_{\rm{EOM}}$ applied to the EOM by the active regulator circuit (R) are recorded versus the angle of incidence $\theta$. The EOM voltage is translated into a phase via the known half wave voltage.}
\label{fig:lock}    
\end{figure}
\section{Results}\label{sec:res}
The operation of the electronic servo lock acting to minimize the power difference $\Delta P$ at the exit ports of the polarization beam splitter in Fig.\ref{fig:setup} is illustrated in Fig.\ref{fig:lock}. In this graph the servo loop was closed and a voltage ramp was applied to the PZT in order to scan the angle of incidence of the laser beam irradiating the Ag film by 14 mrad in 200 ms. The phase of the reflected beam is determined from the voltage fed to the EOM and the known half wave voltage. The plotted error signal is the output of the differential amplifier in Fig.\ref{fig:setup}. When the angle of incidence is swept through the center of the SPR, the required EOM phase changes so rapidly that the bandwidth of the servo lock is not sufficient to maintain the error signal at exact zero. Its deviation from zero, however, is only 2.5$\,\%$ of its total swing for the unlocked case. A larger servo bandwidth could reduce this time lag and thus increase the response time of the SPR sensor. In the present set-up the limited servo bandwidth has constrained us to long sweep times of a few mrad/s in order to maintain the error signal at zero within the noise.
 
\begin{figure}
\center
\resizebox{0.40\textwidth}{!}{
 \includegraphics{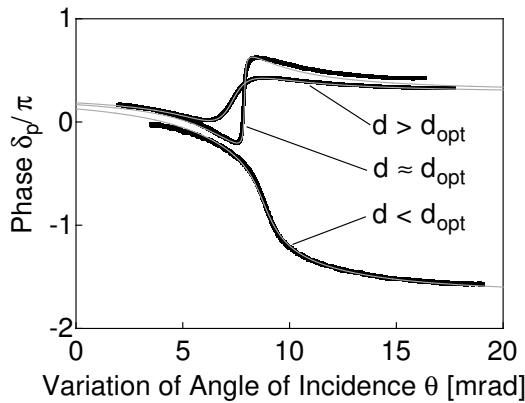}}
\caption{Phase behavior of the SPR for different thicknesses ($d < d_{\rm{opt}}, d \approx d_{\rm{opt}}, d > d_{\rm{opt}}$) of the silver film. The thick black lines show measurements, the thin grey lines are calculated from the Fresnel relations.}
\label{fig:imped}    
\end{figure}

We have used our phase detection technique to record the phase behavior of the SPR for different thicknesses $d$ of the silver film. In the thick black lines in Fig.\ref{fig:imped} the three characteristic cases $d = 48.9 \pm0.7\,$nm$\,> d_{\rm{opt}}$, $d = 45.4\pm1.1\,$nm$\,\approx d_{\rm{opt}}$, $d = 40.5\pm2.5\,$nm$\, < d_{\rm{opt}}$ are illustrated, with $d_{\rm{opt}}$ denoting the thickness corresponding to optimal impedance matching. The thicknesses were determined by taking the mean of the values obtained from fits of the phase data $\delta_p$ and the corresponding reflection data $|r_p|^2$ by means of Fresnel relations. The deviations are indicated as errors. All fits used $\epsilon_{\rm{Ag}}$ and $d$ as free parameters and $\epsilon_{\rm{glass}} = 2.28, \epsilon_{\rm{air}}=1$. The average $\epsilon_{\rm{Ag}}$ for all fits is $\epsilon_{\rm{Ag}} = -(25.2\pm0.2)+i\,(1.6\pm0.2)$.
The fits for $\delta_p$ are shown as the thin grey lines in Fig.\ref{fig:imped}.

\begin{figure}
\center
\resizebox{0.40\textwidth}{!}{
 \includegraphics{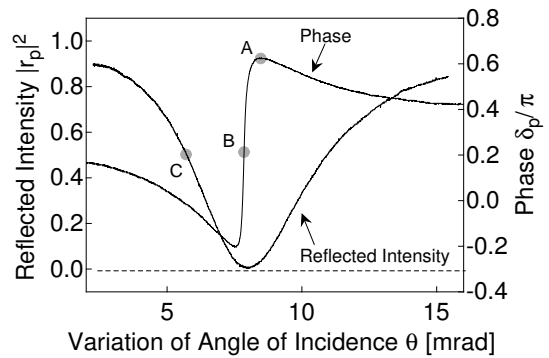}}
\caption{The phase shift $\delta_p$ (measured by recording the phase $\phi_{\rm{EOM}}$ applied via the EOM by means of the closed servo loop) and the intensity reflection coefficient $|r_p|^2$ of the SPR are plotted versus the angle of incidence $\theta$ for nearly optimal film thickness $d \approx d_{\rm{opt}}$. The grey disks denoted by "A" and "B" indicate the angles $\theta$, where the noise analysis in Fig.\ref{fig:noise} has been conducted. The disk denoted by "C" marks the angle $\theta$, where $|r_p|^2$ displays its maximal sensitivity with respect to variations of the refractive index in the vicinity of the Ag film.}
\label{fig:comp}    
\end{figure}

In Fig.\ref{fig:comp} the angle dependence of the phase $\delta_p$ and the intensity reflection coefficient $|r_p|^2$ are compared close to the optimal film thickness $d \approx d_{\rm{opt}}$. Note that $|r_p|^2$ decreases to a minimal value as low as $0.4\%$ in the centre of the SPR. Note also the significantly larger slope of $\delta_p$ in point (B) as compared to that of $|r_p|^2$ in point (C), which reflects the increased sensitivity of the phase signal with respect to changes of the angle of incidence and thus the refractive index adjacent to the Ag film. 

\begin{figure}
\center
\resizebox{0.35\textwidth}{!}{
 \includegraphics{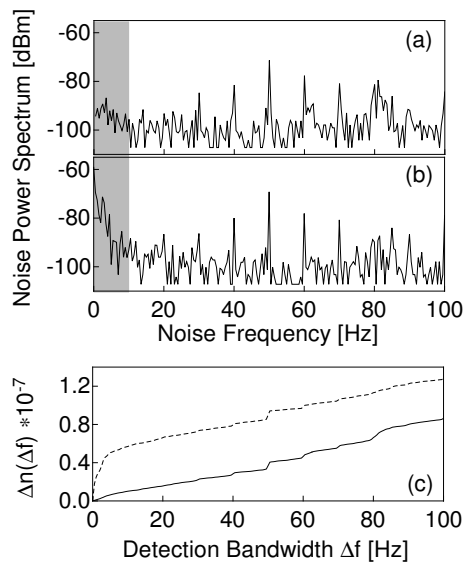}}
\caption{Noise power spectra at the relative maximum (A) and the central turning point (B) of the phase signal in Fig.\ref{fig:comp} are plotted in (a) and (b). The region between 0 and 10 Hz, where both spectra differ, is highlighted by a grey background. In (c) the corresponding refractive-index uncertainties are plotted versus detection bandwidth. The solid line corresponds to (a), the dashed line corresponds to (b).}
\label{fig:noise}    
\end{figure}

In order to quantify the sensitivity of our system with respect to changes of the refractive index while keeping technical noise as low as possible, we have chosen the following procedure: The servo loop is opened and the error signal is nulled by means of a manually adjustable retardation plate, which superseded the EOM. In Fig.\ref{fig:comp} the phase $\delta_p$ was adjusted  to its relative maximum (A) or to its central turning point (B) by setting the appropriate values of $\theta$ with the PZT-mirror using a battery driven voltage source. Applying these settings, we have measured in Fig.\ref{fig:noise} the noise power spectra of the error signal $U_{\rm{err}}$ (output of the differential amplifier in Fig.\ref{fig:setup}). In the case (A) the error signal to first order does not depend on $\delta_p$, i.e., fluctuations $\Delta U_{\rm{err}}$ are mainly resulting from amplifier noise, laser intensity noise (e.g., due to imperfect power balance and unequal spectral response of the photo detectors) and fluctuations $\Delta \theta$ of the angle of incidence, due to technical noise of the PZT voltage source and seismic, acoustic or thermal pointing instabilities. The corresponding noise spectrum in Fig.\ref{fig:noise} (a) is basically constant in the range 0-100 Hz with some spikes at multiples of 10 Hz around the power line frequency of 50 Hz. In the case (B) phase noise $\Delta \delta_p(f)$ at the frequency $f$ directly translates into error signal noise $\Delta U_{\rm{err}}(f)$ according to $\Delta \delta_p(f) = \xi \Delta U_{\rm{err}}(f)$ with $\xi = 3.725$ rad/Volt. This yields the excess noise observed within the frequency range 0-10 Hz (shaded area) in the power spectrum of Fig.\ref{fig:noise} (b). We attribute this phase noise $\Delta \delta_p(f)$ to fluctuations of the refractive index near the Ag film due to air pressure fluctuations in the $10^{-2}$ mBar range. Low frequency pressure fluctuations of this size seem not surprising since the metallic film in our set-up is exposed to the atmosphere in a typical laser laboratory environment with infrasound sources like the air conditioning system in addition to atmospheric sources \cite{Bed:00}. Since phase changes $\Delta \delta_p(f)$ represent the signal of interest here, our detectivity is only limited by the noise observed in Fig.\ref{fig:noise} (a), which has no contribution from changes of $\Delta \delta_p(f)$ since $\partial \delta_p/\partial \theta = 0$ at point (A). We may thus give a conservative estimate of the sensitivity of our SPR sensor as follows: we integrate the noise spectrum of Fig.\ref{fig:noise} (a) over the desired bandwidth $\Delta f$ and calculate a corresponding refractive index uncertainty $\Delta n(\Delta f) = \zeta(B) \, \xi \, \int_{0}^{\Delta f} df \Delta U_{\rm{err}}(f)$ by evaluating the differential quotient $\zeta(B) \equiv (\partial n/\partial \delta_p)_B = 1/16824$ at point (B) in Fig.\ref{fig:comp} (cf. $\zeta(A) = 0.65$). This is shown in the lower trace (solid line) of Fig.\ref{fig:noise} (c). The upper trace (dashed line) shows the result, if instead of the noise spectrum in (a) the spectrum in (b) is used in the same procedure. While the lower trace represents the lower limit for the uncertainty $\Delta n(\Delta f)$ in our experiment resulting from technical deficiencies of our detector set-up, the upper trace adds the extra detection uncertainty due to air pressure fluctuations, which would be absent in a real measurement application. For example, at a detection bandwidth of $\Delta f = 10$ Hz we find $\Delta n(\Delta f) \approx 10^{-8}$ from the solid curve in (c), which compares favorably with previously reported values \cite{Kab:98}. This corresponds to a signal-to-noise ratio in the measured phase signal $\delta_p$ in Fig.\ref{fig:comp} of $1.5 \times 10^{4}$ in 0.1 s observation time, a requirement easily realized with standard electronics. This has to be compared with a standard determination of the reflection coefficient using the left turning point (C) of the reflection signal of Fig.\ref{fig:comp}, where the differential quotient $\partial n/\partial |r_p|^2$ attains its maximal value of about 0.041. Here, a signal-to-noise ratio of about $4.1 \times10^{6}$ in 0.1 s in the determination of $|r_p|^2$ would be required to obtain the same sensitivity with regard to the refractive index $n$. This shows that the phase-shift detection method significantly relaxes the demands with respect to dynamical range and noise figures of the required intensity detectors and electronic processing equipment.

\section{Conclusions}
We have presented a method to directly measure the phase of a laser beam reflected from a metallic film after excitation of surface plasmon polaritons. This method permits real time access to the phase, it increases the possible speed of data acquisition, and it may thus prove useful for increasing the sensitivity of surface plasmon based sensors with reasonable expenses with respect to the detection and data processing equipment. We estimate that in our experiment the index of refraction in a layer adjacent to the metallic film supporting the plasmon resonance could be measured with an uncertainty of about $10^{-8}$ with 10 Hz detection bandwidth. Phase sensitive detection of surface plasmons may also prove useful in the context of atom optical experiments, which exploit the enhanced evanescent wave arising, when a surface plasmon resonance is excited \cite{Ess:93,Gar:07}. 

\section{Acknowledgements}
We are grateful to D. Heitmann and his team for support in the production of the metallic films.

\end{document}